\documentclass[prb,nofootinbib,twocolumn,superscriptaddress]{revtex4} 

\usepackage{graphicx}
\usepackage{dcolumn}
\usepackage{bm}
\usepackage{threeparttable}
\usepackage{times}
\usepackage{mathptmx}
\usepackage{lscape}
\usepackage{natbib}
\usepackage{amsmath}
\usepackage{amssymb}
\usepackage{braket}
\usepackage{comment}
\usepackage{color}

\def\degree{\kern-.2em\r{}\kern-.3em}

\begin{document}


\title{  Optimal Selection of Structural Degree of Freedoms for Spceial Microscopic States to \\ Characterize Disordered Structures  }
   
\author{Koretaka Yuge}
\affiliation{
Department of Materials Science and Engineering,  Kyoto University, Sakyo, Kyoto 606-8501, Japan\\
}%

\author{Shouno Ohta}
\affiliation{
Department of Materials Science and Engineering,  Kyoto University, Sakyo, Kyoto 606-8501, Japan\\
}%

\begin{abstract}
{ For classical discrete systems under constant composition, statistical mechanics tells us that a set of microscopic state dominantly contributing to thermodynamically equilibrium state should depend on temperature as well as on many-body interaction (i.e. thermodynamic information), through Boltzamann factor of $\exp\left( -\beta E \right)$. Despite this fact, our recent study reveals that a single (and a few additional) microscopic state (called projection state: PS), whose structure can be known \textit{a priori} without requiring thermodynamic information, can universally characterize equiibrium properties for disordered states, where their sturctures depends on configurational geometry \textit{before} applying many-body interaction to the system. Although mathematical condition for the structures of PS have been rigorously established, practically effective condition for constructing the stuructures, especially for which set of a \textit{finite} structural degree of freedoms (SDF) should be selected for considered coordination has not been clarified so far. We here tuckle this problem, proposing a quantitative and systematic criteria for an optimal set of SDFs. The present proposal enables to effectively constructing PSs for a limited system size, and also providing new insight into which set of SDFs should generally affects equilibrium properties along a chosen coordination, without using any thermodynamic information. 

  }
\end{abstract}


\maketitle

\section{Introduction}
When we consider structure along coordination of $Q_{r}$ out of prepared set of coordination $\left\{ Q_{1},\cdots, Q_{f} \right\}$ under thermodynamically equilibrium state for classical discrete systems with constant composition, it can be typically given by the so-called canonical average of 
\begin{eqnarray}
\label{eq:can}
\Braket{Q_{r}}_{Z} = Z^{-1} \sum_{d} Q_{r}^{\left( d \right)} \exp\left( -\beta E^{\left( d \right)} \right).
\end{eqnarray}
Here, $Z$ denotes partition function, $\beta$ inverse temperature, and summation should be taken over all possible microscopic states on configuration space (i.e., trace over possible configurations). 
Canoical average of Eq.~\eqref{eq:can} certainly tells that a set of microscopic state dominantly contributing to $\Braket{Q_{r}}_{Z}$ should in principle depend on temperature as wll as on energy, which therefore cannot be known \textit{a priori} without any thermodynamic information.
WIth these considerations, a variety of theoretical approaches has been developed to efficiently sample important states such as Metropolis algorism, entropic sampling and Wang-Landau method.\cite{mc1,mc2,mc3,wl} 
Our recent study reveal that despite these facts, by using information about configurational geometry \textit{before} applying many-body interaction to the system, we can \textit{a priori} know a set of important microscopic state to characterize disordered structure, without using any thermodynamic information. In this context, canonical average of Eq.~\eqref{eq:can} can be simplyfied to\cite{em1,em2,em0}
\begin{eqnarray}
\label{eq:emrs}
\Braket{Q_{r}}_{Z} \simeq \Braket{Q_{r}} -  C_{r}\beta E_{r} + \frac{\beta^{2}}{2}\sum_{i=1}^{g} \omega_{i} E_{r_{i}}^{2},
\end{eqnarray}
where $\Braket{\quad}$ denotes taking arithmetic average over configurational density of states (CDOS), $C_{r}$ represents constant depending only CDOS geometry, and $E_{r}$ and $E_{r_{i}}$s are energy for the special microscopic states (respectively called as PS and PS2s), whose structures can be known \textit{a priori} without any thermodynamic information, depending only configurational geometry. 
In our previous study, although mathematical condition for the structures of PS and PS2 have been rigorously established, practical condition for effectively constructing special microscopic states, especially for finding optimal set of SDFs for a chosen coordination, has not been clarified so far. This should be practically, significantly important to construct special atomic configurations under limited system size, as well as imporntant to know which set of SDF affects equilibrium properties along chosen coordination.
We here theoretically tuckle this problem using information only about configurational geometry, and establishing a quantitative and systematic criteria for selecting optimal set of SDFs for a chosen coordination. The details are shown below. 


\section{Derivation and Discussions}
\begin{figure}[h]
\begin{center}
\includegraphics[width=1.0\linewidth]{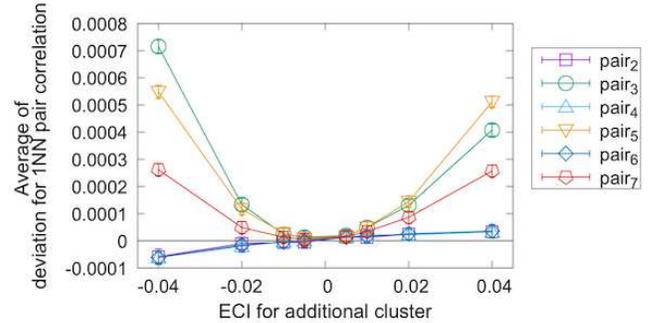}
\caption{Comparison of effects of another SDF to the considered SDF of 1NN pair, when effective interactions for another SDF takes finite value while the considered interaction takes zero.}
\label{fig:eci}
\end{center}
\end{figure}

To pracitcally construct atomic configurations for special microscopic states with given set of known multisite correlations for alloys, we should typically, firstly include possible pair correlations, and additional higher-order correlations. Especially for multicomponent (number of components is greater than two) alloys, number of basis functions for multisite correlations should exponentially increase with increase of dimension of figure, we here confine ourselves to consider a selection of optimal set of pair coordinations. 
Then, we here consider $f$-degree of freedom system (i.e., up to $f$-th nearest neighbor ($f$-NN) pair correlation) on given periodic lattice under constant composition, and focusing on canonical average for $Q_{r}$. 
Under these preparations, $j$-th coordination for PS sturucture is explicitly given by
\begin{eqnarray}
Q_{j}^{\left( r \right)} \simeq \sqrt{\frac{2}{\pi }}\Braket{Q_{r}}_{2}^{-1} \mu_{2}^{\left( r,j \right)}, 
\end{eqnarray}
where $\Braket{\quad}_{2}$ denotes taking arithmetic average for configurational density of states (CDOS), and $m_{2}^{\left( rj \right)}$ denotes $\left( r,j \right)$-component of 2-order moment matrix for the CDOS.  Since we have previously found through exact formulation of moments in CDOS that 
\begin{eqnarray}
\lim_{N\to\infty}\mu_{2}^{\left( rj \right)} = \delta_{rj},
\end{eqnarray}
it should be practically difficult to optimize and decrease number of SDF for PS structure.
Meanwhile, $j$-th component of $i$-th structure of PS2 along $r$-th coordination is given by
\begin{eqnarray}
\label{eq:3mom}
Q_{j}^{\left( {r_{i}} \right)}  = \lambda_{i}^{\frac{1}{2}} U_{ij},
\end{eqnarray}
where $\lambda_{i}$ and $U_{ij}$ are $i$-th singular value and $\left( i,j \right)$ component of left singular vector for real symmetric $f\times f$ 3-order moment matrix $\mathbf{A}$ defined as $A_{pq} = \Braket{Q_{r}Q_{p}Q_{q}} =\mu_{3}^{\left( r,p,q \right)}$.
Then, our strategy is to find a optimal set of e.g., $\left( f-m \right)$ out of $f$ SDFs, to provide the best approximation for the 3rd term of r.h.s. in Eq.~\eqref{eq:emrs} over all possible many-body interaction. 
To achieve this, we here employ generalized Ising model (GIM) to exactly describe potential energy for any given configuration $s$:
\begin{eqnarray}
U_{s} = \sum_{i=1}^{f} \Braket{U|Q_{i} }Q_{i}^{\left( s \right)},
\end{eqnarray}
where $\Braket{\quad|\quad}$ denotes inner product, i.e., trace over configuration space. 
Since the third term of Eq.~\eqref{eq:emrs} can be described by the quadratic form for matrix $\mathbf{A}$ and the inner products, we should naturally minimize the followings on $S^{f-1}$ hypersphere integral satisfying $\sqrt{\sum_{i=1}^{f}\Braket{U|Q_{i}}^{2}}=1$:
\begin{widetext}
\begin{eqnarray}
\label{eq:R}
G &=&  \sqrt{ \int_{S^{f-1}} \left\{ \sum_{j,k} \Braket{q_{r}q_{j}q_{k}} \Braket{E|q_{j}}\Braket{E|q_{k}}  -    \sum_{j\notin \left\{ m \right\},k\notin \left\{ m \right\}} \Braket{q_{r}q_{j}q_{k}} \Braket{E|q_{j}}\Braket{E|q_{k}}   \right\}^{2} d\mathbf{X} \bigg/ \int_{S^{f-1}} d\mathbf{X}}        \nonumber \\
&=& \sqrt{\frac{  \left( \sum_{i\in\left\{ m \right\}} a_{ii} \right)^{2} + 2 \sum_{i\in\left\{ m \right\},j\in\left\{ m \right\}} \left( a_{ij} \right)^{2}  }{f\left( f+2 \right) }},
\end{eqnarray}
\end{widetext}
where $a_{jk}$ is $\left( j,k \right)$ element of $\mathbf{A}$.
We have previously shown that at thermodynamic limit, element of $\mathbf{A}$ for A$_{x}$B$_{\left( 1-x \right)}$ binary system can be exactly given by 
\begin{eqnarray}
\label{eq:a}
\lim_{N\to\infty}a_{jk} = \frac{64z^{2}}{N^{2}D_{r}D_{j}D_{k} }\left\{ \left( 1-4z \right)D_{r}\left[ r=j=k \right]+ z c_{rjk} \right\},
\end{eqnarray}
where $N$ denotes number of lattice points in the system, $z=x\left( 1-x \right)$, $D_{r}$ is the number of pair $r$ per site, $c_{rjk}$ is the number of closed path consisting of $r,j,k$ pairs, and $\left[ \quad \right]$ represents Iverson braket.
At equiatomic composition, especially, $a_{jk}\propto c_{rjk}/\left( D_{j}D_{k} \right)$, intuitive strategy to find optimal set of SDF is to consider a set of SDF with smaller number of closed path including $r$-th pair, and larger number of the rest two pairs. 
Figure~\ref{fig:eci} shows example of the comparison of effects of another SDF to the considered SDF of 1NN pair, when effective interactions for another SDF takes finite value while the considered interaction takes zero. We can clearly see that effects of interaction for another SDF to the considered SDF strongly depends on the choice of the SDF, which should be systematically and geometrically interpreted according to the proposed formulation of 
Eqs.~\eqref{eq:R} and~\eqref{eq:a}. 


\section{Conclusions}
We here theoretically examine a optimal set of structural degree of freedom (SDF) for special microscopic states characterizing equilibrium properties, using information only about configurational geometry. By performing hypersphere integration, we find that such set can be intuitively obtained by omitting a set of SDF with lower number of corresponding triplet closed path and higher number of pairs other than the considered coordination. Further quantitative relationships between an optimal set of SDF including higher-dimension figure and multicomponent systems should be clarified in our future works.

\section{Acknowledgement}
This work was supported by Grant-in-Aids for Scientific Research on Innovative Areas on High Entropy Alloys through the grant number JP18H05453 and a Grant-in-Aid for Scientific Research (16K06704) from the MEXT of Japan, Research Grant from Hitachi Metals$\cdot$Materials Science Foundation, and Advanced Low Carbon Technology Research and Development Program of the Japan Science and Technology Agency (JST).

\end{document}